\documentclass[10pt,conference]{IEEEtran}
\makeatletter
\def\ps@headings{%
\def\@oddhead{\mbox{}\scriptsize\rightmark \hfil \thepage}%
\def\@evenhead{\scriptsize\thepage \hfil \leftmark\mbox{}}%
\def\@oddfoot{}%
\def\@evenfoot{}}
\makeatother
\IEEEoverridecommandlockouts
\usepackage{cite}
\usepackage{amsmath,amssymb,amsfonts}
\usepackage{algorithmic}
\usepackage{graphicx}
\usepackage{textcomp}
\usepackage{comment}
\usepackage[T1]{fontenc}
\usepackage{longtable}
\usepackage{lscape}
\usepackage{caption}
\usepackage{subcaption}
\usepackage{multirow}
\usepackage{graphicx}
\usepackage[table,xcdraw]{xcolor}
\usepackage[normalem]{ulem}

\def\BibTeX{{\rm B\kern-.05em{\sc i\kern-.025em b}\kern-.08em
    T\kern-.1667em\lower.7ex\hbox{E}\kern-.125emX}}

\definecolor{darkgreen}{rgb}{0.29, 0.33, 0.13}
\definecolor{std_red}{rgb}{1, 0, 0}
\definecolor{std_green}{rgb}{0, 1, 0}

\newcommand{\system}{XRP-NDN Overlay}
\begin{document}
\bstctlcite{IEEEexample:BSTcontrol}

\title{\system : Improving the Communication Efficiency of Consensus-Validation based blockchains with an NDN overlay 
}

\author{
\IEEEauthorblockN{
Lucian Trestioreanu\IEEEauthorrefmark{1},
Wazen M. Shbair\IEEEauthorrefmark{1},
Flaviene Scheidt de Cristo\IEEEauthorrefmark{1},
and
Radu State\IEEEauthorrefmark{1}
}
 \IEEEauthorblockA{\IEEEauthorrefmark{1} University of Luxembourg, SnT, 29, Avenue J.F Kennedy, L-1855 Luxembourg\\
 Email:\{flaviene.scheidt, wazen.shbair, lucian.trestioreanu, radu.state\}@uni.lu\\
 }
 }

\maketitle


\begin{abstract} 

With the growing adoption of Distributed Ledger Technologies and the subsequent scaling of these networks, there is an inherent need for efficient and resilient communication used by the underlying consensus and replication mechanisms. While resilient and efficient communication is one of the main pillars of an efficient blockchain network as a whole, the Distributed Ledger Technology is still relatively new and the task of scaling these networks has come with its own challenges towards ensuring these goals. New content distribution concepts like Information Centric Networking, of which Named Data Networking is a worthy example, create new possibilities towards achieving this goal, through in-network caching or built-in native multicasting, for example.

We present and evaluate \system, a solution for increasing the communication efficiency for consensus-validation based blockchains like the XRP Ledger~\cite{DBLP:journals/corr/abs-1802-07242}. We experiment by sending XRP Ledger's consensus messages over different Named Data Networking communication models and prove that our chosen model lowers the number of messages at node level to minimum necessary, while maintaining or improving blockchain performance by leveraging the possibilities offered by an overlay such as specific communication mechanisms.

\end{abstract} 



\section{Introduction}
\label{sec:intro}

While different aspects concerning Distributed Ledger Technology (DLT) research have lately benefited from increased attention from the community, the underlying communication mechanisms, often relying on flooding mechanisms due to the one-to-many and many-to-many communication needs of DLT, have received somewhat less attention. The task of scaling the blockchain networks while maintaining their performance and resilience comes with its own specific challenges, one of these being to maintain or even improve the efficiency and resilience of the underlying communication when working at scale. Each blockchain technology type has its specifics, and per our current understanding, a one-size-fits-all solution is far from possible. For example in the case of Ethereum (ETH), which until recently was a Proof of Work (PoW) blockchain, Gossipsub~\cite{gossipsub-paper} was proposed to improve its communication layer, while \cite{NDN-Ethereum} have proposed a Named Data Networking (NDN)-based design for the ETH block propagation. However the community effort was mainly directed towards PoW-type DLTs, with other types like consensus-validation based blockchains receiving less attention. The XRP Ledger (XRPL) is such a consensus-validation based DLT which has a high throughput, low fees, it is fast, and has a low energy consumption footprint \cite{DBLP:journals/corr/abs-1802-07242},\cite{RPCA-old}. Here, the size of the messages involved in the consensus protocol, e.g. validations and proposals is small enough (approximately 0.5kB) as to not pose a challenge regarding message size in itself. Instead, in the context of the near-real-time communication needs (a new ledger is created every 3 to 5 seconds), it is rather the overwhelming number of messages and the processing incurred at each node, inherent to the flooding model of communication used in XRPL, that finally challenges the scalability goals of the XRPL network.


At scale, the overhead incurred by the flooding of messages increases the requirements for the communication channels e.g. bandwidth (BW), the hardware used by the nodes (CPU and memory), and the energy and financial burden, which if not addressed, could finally result in a degradation of the overall network performance. 

To mitigate the flooding overhead, different approaches to improve the message dissemination efficiency could be considered, e.g. improving the efficiency of the dissemination protocol itself, or external solutions such as overlays.

As such, in the case of XRPL, the problem can be stated, in a broad manner: How coan the performance burden added to the nodes' CPUs, memory, and bandwidth by the high number of messages incurred by the flooding dissemination model used at scale, could be alleviated? Of the possible solutions, we focus on improving the dissemination method to decrease the number of messages, and deviating (some of) these messages through an overlay where we can make use of specific properties to achieve this goal.

NDN~\cite{Brief_Intro_NDN} is a proposal for a Future Internet Architecture which instead of delivering packets to a given destination (IP), fetches the data by name, offering for example \textit{content caching} to improve delivery speed and reduce congestion. The process of DLT data dissemination can also benefit from native, built-in \textit{multicast} available on NDN.

Data can be disseminated on NDN in two manners:

1) In the classic, native pull-based approach, interested nodes request and receive by name the pieces of data that they are interested in; this, combined with NDN in-network caching, could increase communication efficiency by decreasing the overall number of messages exchanged.

2) In another approach, also taken by \cite{BoNDN} for example on DLT, the data can be encapsulated in the Interest Packet and disseminated with multicast on pre-determined paths (interested nodes must enable multi-cast for the respective Interest). Moreover, current work to decrease the number of duplicate messages on NDN multicasting \cite{multicast_duplicate_suppression}, could further improve communication efficiency of XRPL.

Our contribution is two-fold: i) to our knowledge there has been no prior work on this topic concerning consensus-validation based DLTs, and ii) we evaluate over multiple practical implementation models, to find the best approach that benefits a concrete case such as the XRPL.


\section{Background}
\label{sec:bk}

\subsection{The XRP Ledger}

The XRP Ledger, one of the most established DLTs to date, is characterised as an open-source, permissionless and decentralized blockchain system. It is appreciated for its transaction (tx) throughput (currently +/- 1500 tx/s), speed (can settle a tx in 3-5s) and low fees (in the order of \$0.0000774 for example as of April 2021). It is also considered eco-friendly, with a low energy consumption mainly due to the fact that by design does not make use of, for example, Proof of Work. 



The two main phases of XRP's blockchain building process are called \textit{"Consensus"} and \textit{"Validation"}. During \textit{Consensus}, new transactions are received and the participants agree on: the transaction set to apply over the previous agreed-upon ledger AND on the closing time of the newly created ledger. During validation, participants agree on which ledger version was generated, based on the ledgers created by chosen peers. 

The XRPL consensus is generally classified as Byzantine fault tolerant, however nodes do not know all the other participating nodes, and consensus wise a node only communicates with those from its own Unique Node List (UNL), where UNL is defined as a set of nodes that an individual node does not necessarily consider to be all honest, but instead it trusts not to collude to fraud.

The main phases of the \textit{Consensus} are:

\begin{enumerate}
    \item \textit{"Open" phase} - New tx's are received. To reach as many nodes as fast as possible, a flooding mechanism is currently being used for tx propagation.
    \item \textit{"Close" state} - The current ledger won't accept new tx's, instead the \textit{Consensus} protocol advances towards closing the current ledger. Tx's received after this moment will be recorded and applied to next ledger.
    \item \textit{"Establish" phase} - The participants work towards agreeing on the current tx set by exchanging \textit{proposal} messages, adding or removing tx's. They also agree on the effective close time. In this phase, flooding is also being used for the propagation of the proposal messages.
    \item \textit{"Consensus reached" state} - Participants agreed on the tx set to include in the current ledger.
    \item \textit{"Accept" phase} - Participants apply the agreed upon tx set in canonical order and share the result.
    \item \textit{"End round" state} - the current round is finished, and the participants move to \textit{Validate} this ledger.
\end{enumerate}

During \textit{Validation}, the validator nodes share their results as signed messages containing the hash of the calculated ledger. These messages are called \textit{validations} and allow participants to check if they obtained identical results; then, they can declare the ledger "final". The propagation of validation messages also uses the flooding mechanism. Validators compare their results and declare the ledger validated IF enough trusted validators agree. This number of nodes is also called "quorum".

Authors of \cite{security_analysis_xrpl} identified relatively simple cases where consensus may violate safety and/or liveness; it is argued that the XRPL needs a very close synchronisation, interconnection and fault free operation between validators.

Flooding is the main dissemination protocol on XRPL, an approach which ensures robustness. The main types of flooded data are \textit{Transactions} (Tx), \textit{Proposals}, and \textit{Validations}.

\subsection{Named Data Networking (NDN)}

Named Data Networking (NDN) is a proposed \textit{Future Internet} architecture evolved from the 2006 \textit{Content-Centric Networking} (CCN) project, by Van Jacobson. 

Noticing that today's Internet is rather used as an information distribution network, NDN is not delivering packets to a given destination address (IP), but it fetches the data which is identified by a given name. NDN distinguishes itself from other architectures through
the following:
\begin{enumerate}
    \item In NDN, the \textit{data} is named by the applications, and \textit{Consumers} request data by its \textit{name}. As such the process is consumer-driven.
    \item \textit{Data packets} are cryptographically signed by their respective \textit{Producers}. As a result of this \textit{data-centric} approach, the data can be verified by consumers no matter how it was received.
    \item Routers record each data request \textit{(Interest packet)} and erase it once data is received. As such, smart strategies can be used for forwarding, and loops eliminated.
\end{enumerate}

NDN offers \textit{content caching} to improve delivery speed and reduce congestion, a \textit{simpler configuration} of the network devices, and \textit{data-level security}.

\begin{figure}[ht]
\begin{center}
    \includegraphics[width=0.485\textwidth]{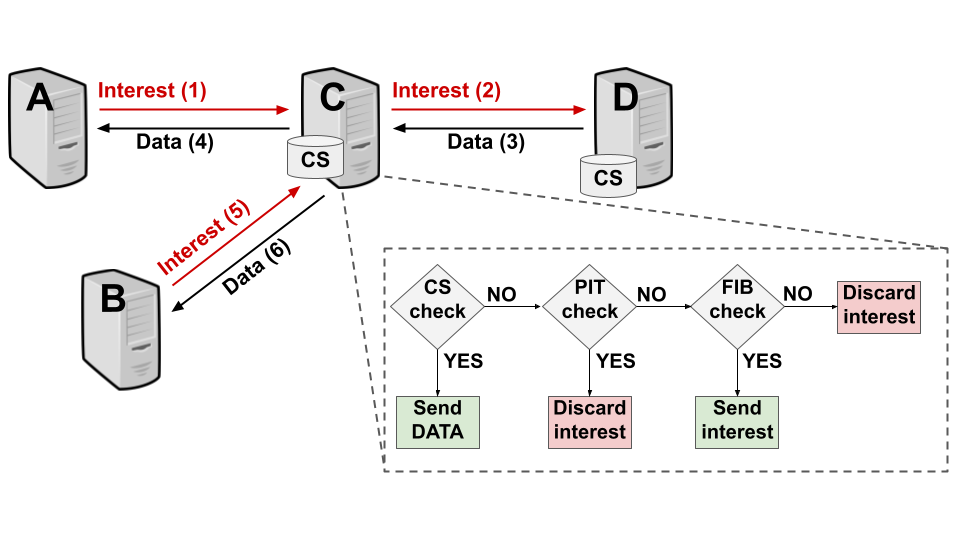}
    \caption{The basic mechanisms of NDN}
    \label{fig:NDN-model}
\end{center}
\end{figure}

On NDN, the \textit{Producer} creates new data, and the \textit{Consumer} is \textit{interested} to receive or \textit{"consume"} the new data produced by the \textit{Producer}. From these two roles derive the packet types:
\begin{enumerate}
    \item The \textit{Interest packet}, normally sent by a \textit{Consumer} to ask for some data piece produced by a \textit{Producer}.
    \item The \textit{Data packet}, which is normally created by a \textit{Producer} and sent back to a \textit{Consumer} as a response to an \textit{Interest} packet sent by the \textit{Consumer.} 
\end{enumerate}

Besides the above, other relevant NDN building blocks are:

\begin{enumerate}
    \item \textit{Content Store (CS)} which stores for some period the data packets which it has already seen in order to immediately serve them in case of a new request.
    \item The \textit{Pending Interest Table (PIT)} is a storage for unfulfilled Interest packets.
    \item The \textit{Forward Information Base (FIB)}, similar to a routing table, helps an NDN node decide where and how to route some packet.
\end{enumerate}

As illustrated in Figure~\ref{fig:NDN-model}, NDN is working as follows:

1) The \textit{Consumer} node A is interested to receive some piece of \textit{Data} and sends the \textit{Interest(1)}.

2) Upon receiving the interest, the \textit{Router} node C:

    - Checks its own \textit{Content Store} to see if \textit{data} is already available locally. If data is available it answers with the data, or in case the data is not available locally it proceeds to next check. In the Figure, data is not available locally, so C proceeds to the next check:
    
    - It looks in its own \textit{Pending Interest Table} (PIT) to see if a similar interest was already received. If a similar interest is already received, the router discards the newly received interest, or if it can't find an entry (this is a new interest), it then proceeds to the next check. In Figure~\ref{fig:NDN-model}, this is a fresh new Interest, so C proceeds to the next check:
    
    - It checks its \textit{Forwarding Information Base (FIB)} to see if it knows how to route the interest further. If it does, it forwards the Interest, or if it can't find an entry it drops it. In our Figure, C has a routing entry in its FIB so it forwards the \textit{Interest(2)} to \textit{Producer} node D.
    
    - Upon receiving the Interest, D answers by sending the \textit{Data(3)} which then is stored by node C in its CS, AND forwarded to node A as the \textit{Data(4)} packet.
    
3) When \textit{Consumer} node B sends the \textit{Interest(5)} for same \textit{Data}, the data is found by C in its CS, and is directly forwarded to B as the \textit{Data(6)} packet in Figure~\ref{fig:NDN-model}.


\section{Design and Implementation}
\label{sec:design}

In this section we describe our solution \system~for improving the communication efficiency in the case of the blockchains that use a consensus-validation based system, working on the concrete case of the XRP ledger.

We chose NDN as overlay because it offers: i) in-network caching of data, which on large networks  can lower the overall number of messages in-flight at a given moment, and ii) native multicasting, which could also soon benefit from mechanisms of reducing message duplicates \cite{multicast_duplicate_suppression}.


With a goal to decrease the load on the XRPL nodes by decreasing the number of messages processed by them, we seek to answer the following questions:

\begin{enumerate}
    \item[\textbf{Q1}] What models could we use to map the XRPL consensus protocol to the NDN communication environment?
    \item[\textbf{Q2}] How would the models considered compare between each other and with the baseline (unmodified XRPL communication)?
\end{enumerate}

\begin{figure}[ht]
\begin{center}
    \includegraphics[width=0.485\textwidth]{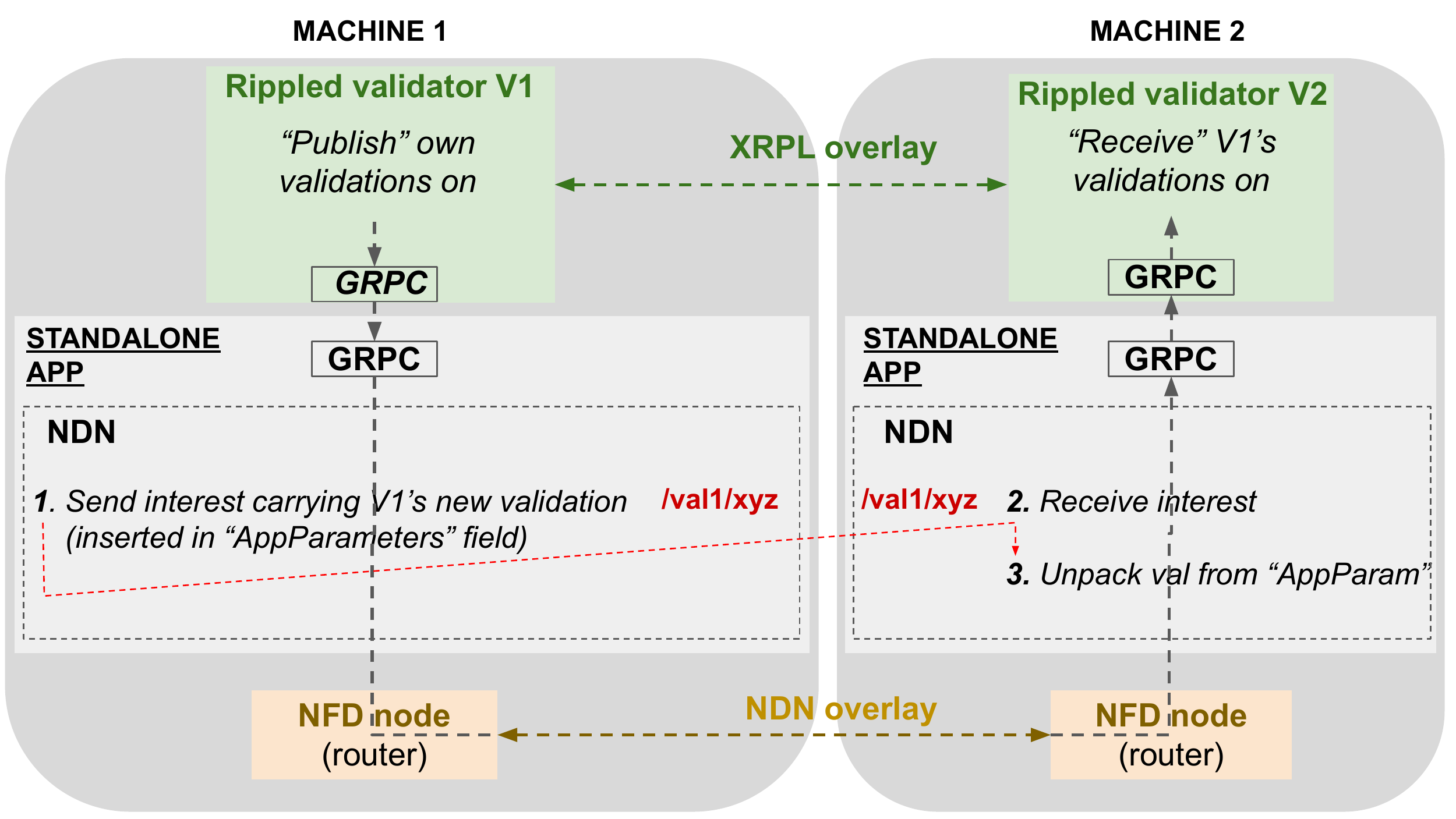}
    \caption{The \textit{piggyback} model}
    \label{fig:piggyback model}
\end{center}
\end{figure}

\begin{figure}[ht]
\begin{center}
    \includegraphics[width=0.485\textwidth]{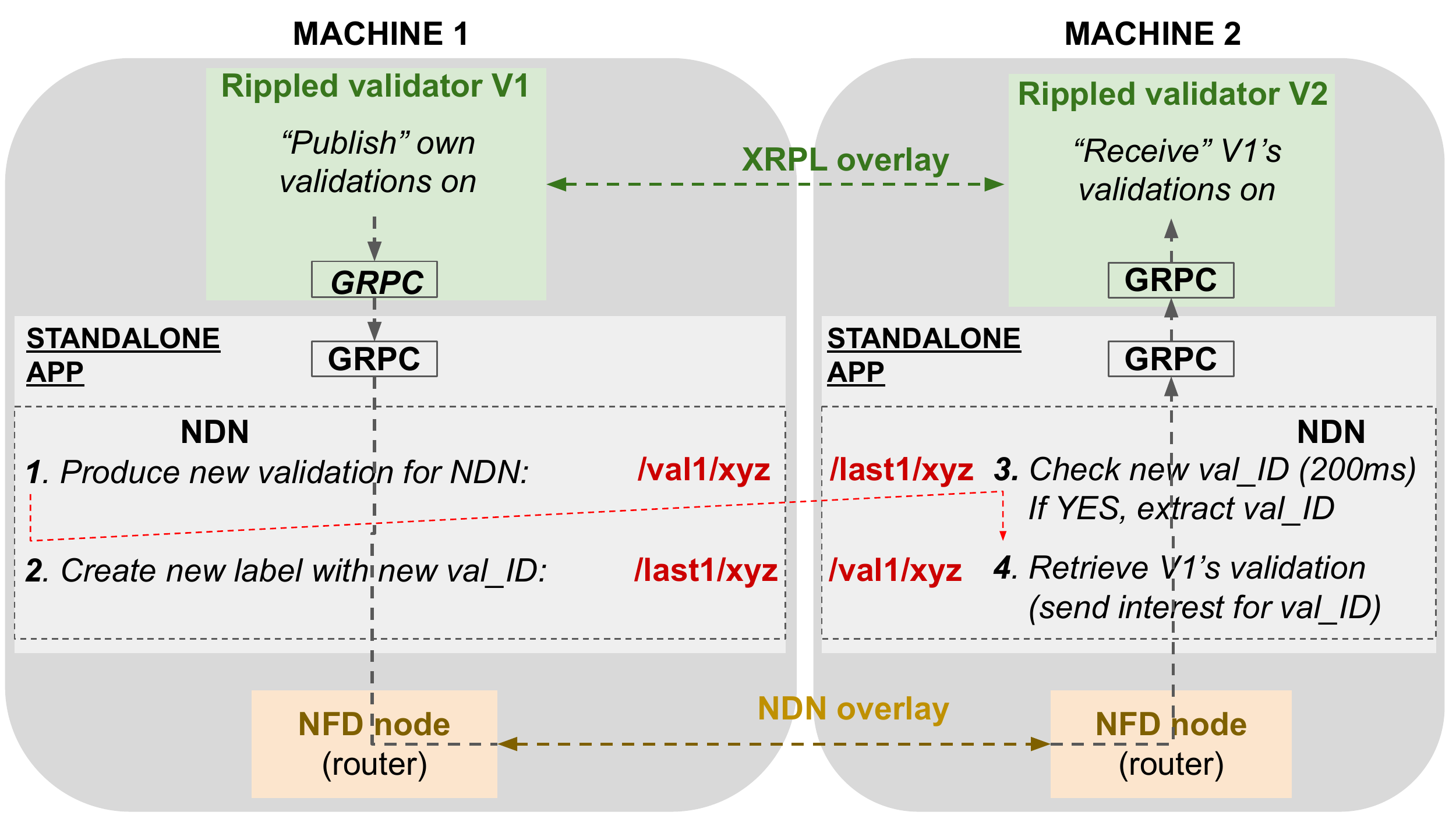}
    \caption{The \textit{polling} model}
    \label{fig:polling model}
\end{center}
\end{figure}

Concerning Q1, we evaluated three possible models for sending  XRPL validations over NDN:

\begin{enumerate}
    \item \textit{"Polling"}: Each validator maintains a \textit{"sequence\_number"}, i.e., to each newly created validation it associates an increasing \textit{"sequence\_number"}. The nodes interested to receive a validation from this validator, will send periodic interests asking \textit{"what is the last \textit{sequence} of your validations?"}. If the sequence is unchanged, they do nothing, and if the sequence increased, they ask for the new validation. As the interval between ledgers on XRPL is normally 3-5 seconds, we chose a 200ms polling interval to ensure we don't delay much the propagation of any fresh validation. The process is illustrated in Figure~\ref{fig:polling model}.
    
    
    \item \textit{"Announce-pull"}: A validator which has created a new validation, would send a multicast interest to let all nodes know the new sequence of its new validation. The interested nodes will pull the validation with the given sequence. The process is illustrated in Figure~\ref{fig:announce-pull model}.

    \item \textit{"Advanced-request"}:  On XRPL, because a \textit{Consumer} knows in advance the identity of the originating \textit{Producer} (a validator on its UNL), and because the interval between validations is somewhat predictable (3-5s in real-life), it is possible to consider the announcement of a new validation made even before the validation is produced. Thus, the time required to forward the interest to source can be eliminated by proactively requesting the validation in advance, as illustrated in Figure~\ref{fig:advance-request model}. This  pull-based approach can ensure that the data is served as soon as it is available.

\begin{figure}[ht]
\begin{center}
    \includegraphics[width=0.485\textwidth]{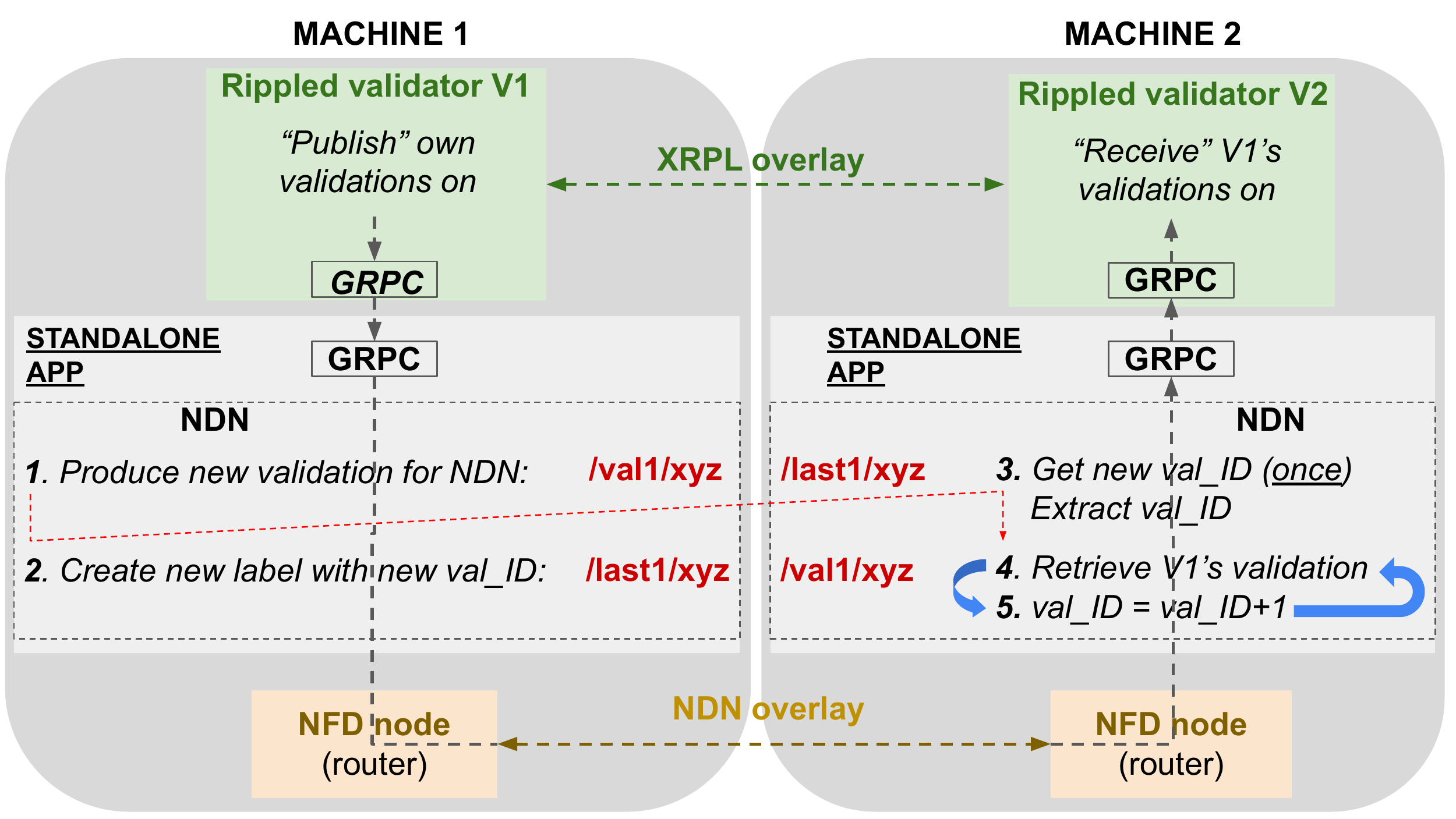}
    \caption{The \textit{advance-request} model}
    \label{fig:advance-request model}
\end{center}
\end{figure}    

\begin{figure}[ht]
\begin{center}
    \includegraphics[width=0.485\textwidth]{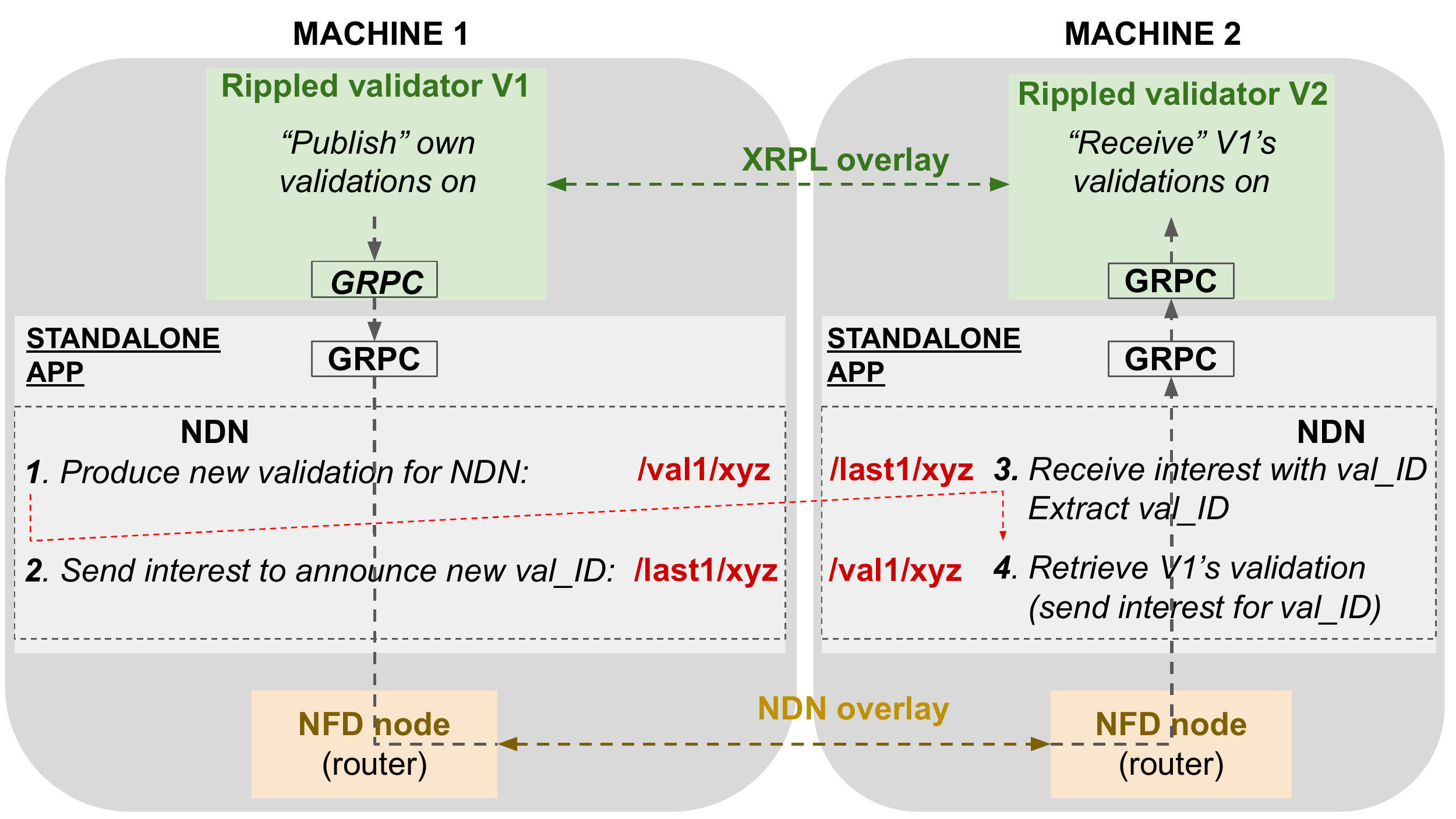}
    \caption{The \textit{announce-pull} model}
    \label{fig:announce-pull model}
\end{center}
\end{figure}
    
    
    \item \textit{"Piggybacking on Interest"}: We notice that it is feasible to send the XRPL validations directly over NDN Interests by encapsulating them in the field named \textit{appParameters} from the Interest packet. Normally this field is meant to carry custom extra data which could eventually be necessary to disambiguate, or help define completely, the request expressed by the Interest message, and the data format is similar to the \textit{Data Packet} response issued by the content \textit{Producer}. In this model, the validator which has created a new validation will encapsulate it in an Interest message and send it directly with multicast to all nodes. This work-around can help reduce at minimum possible the number of messages exchanged at the NDN overlay level because for the dissemination of a validation, only one message (the Interest) is sent, instead of the regular exchanges \textit{Consumer-Producer}. This approach can also help latency-wise in some cases (no two way request-response here), because the data caching available in the pull model can help latency-wise especially when multiple nodes request same data on a same data path. The process is illustrated in Figure~\ref{fig:piggyback model}.
\end{enumerate}

To answer Q2, we looked at the following aspects:

\begin{enumerate}
  \item[\textbf{M1}] \textit{XRPL node load}: How do our NDN models compare with each other and with the baseline concerning the number of validations in/out at node level?
  \item[\textbf{M2}] \textit{Network load}: How do our NDN models compare with each other and with baseline concerning total number of messages and bytes required to travel the overlay in order to propagate a validation to all nodes? 
  \item[\textbf{M3}] \textit{XRPL network stability}: How do our NDN models affect the inter-arrival times of the validation messages?
  \end{enumerate}
  
Besides our proposed models, we also analysed:
\begin{enumerate}
    \item The validation messages for UNL validators in the live XRPL network, presented below. This gave us information about the behavior of XRPL messaging in real-life.
    \item The same for a private network of unmodified XRPL validators fully connected in-between each other, which is our \textit{baseline}.
\end{enumerate}





\section{Experimental Results}
\label{sec:results}

To conduct our evaluation we used both an internal testbed as well as the XRPL live production network. The version of XRPL was 1.7 for the baseline, and we modified it to be able to divert the validation messages through gRPC towards our NDN overlay. Concerning NDN, we used the \textit{NDNts} typescript library~\cite{ndnts-git},\cite{ndnts-home}.

\begin{figure}[ht]
\begin{center}
    \includegraphics[width=0.5\textwidth]{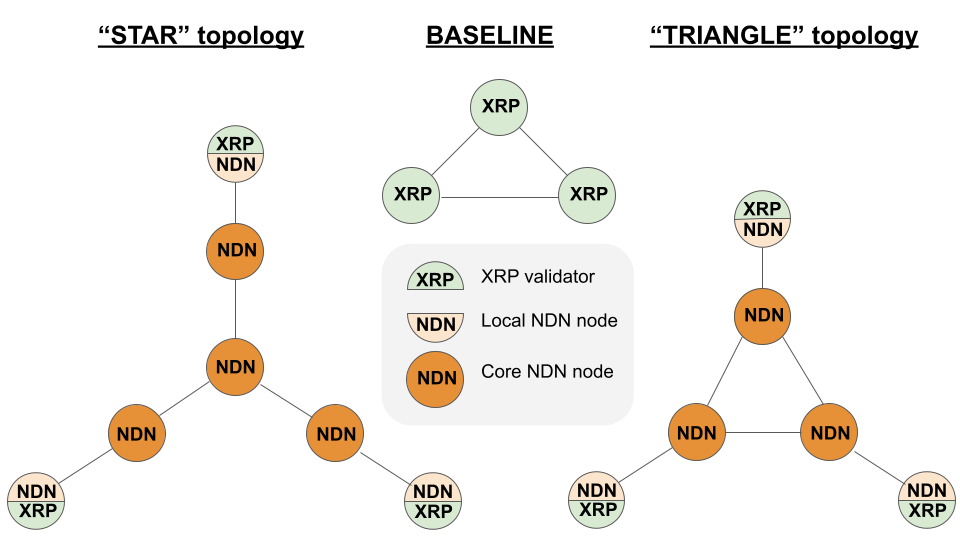}
    \caption{The experimental topologies}
    \label{fig:experimental-topologies}
\end{center}
\end{figure}

The experiments were performed on the below three topologies, also illustrated in Figure~\ref{fig:experimental-topologies}.

\begin{enumerate}
    \item \textit{Baseline} - 3 unmodified XRPL nodes fully connected.
    \item \textit{Star} - 7 NDN nodes linked in a star formation, of which the 3 nodes at the edges are also XRPL nodes.
    \item \textit{Triangle (tri)} - 6 NDN nodes linked in a triangle formation. The 3 edge nodes are also XRPL nodes.
\end{enumerate}

\textbf{M1}: For the unmodified XRPL, we used \textit{Rippled Monitor}\footnote{https://github.com/ripple/rippledmon} and \textit{Grafana} to collect statistics concerning the total number of validations coming in and going out from a node, as well as the total bytes incurred by these messages. For our modified XRPL working over NDN, we counted the number of validations in/out of the node with our own tool.

\textbf{M2}: We used \textit{vnstat} and \textit{tshark} to count the number of bytes/packets at the local machine NIC level.

\textbf{M3}: We parsed specific lines in the XRPL log and extracted the necessary info to analyze the inter-arrival times per validator. To facilitate the readability of our time-series figures, we plotted: 

- in \textit{orange color}, the \textit{rolling mean} (\(rm\)) over the previous 20 data-points (which means generally over 1-2 minutes depending on the interarrival times).

- in \textit{green color}, the \(rm(20)\) plus 2 times the \textit{rolling standard deviation} (\(rSTD)\) computed over the same 20 data-points: \(rm(20) + 2*rSTD(20)\).

- in \textit{red color}, the same \(rm(20)\) from which we substract 2 times the \(rSTD(20)\), i.e.: \(rm(20) - 2*rSTD(20)\).

\subsection*{Results}

\subsubsection{Production validators} We deployed and connected an XRPL validator node on the live XRPL network. From this node, we listen for incoming validations from each of the approximately 35 XRPL validators on the official UNL. We record only the first received validation from each of these \textit{trusted validators}, and drop the duplicate messages. 

\textbf{M1, M2}: We didn't collect any data because in the case of the livenet, it is not possible to perform a fair comparison with our models using these metrics. The main reasons are the number of nodes involved, the topology and the real-life internet environment that we can not recreate for our models.

\begin{figure}[t!]
    \centering
    \begin{subfigure}[t]{0.5\columnwidth}
        \centering
        \includegraphics[height=1.15in]{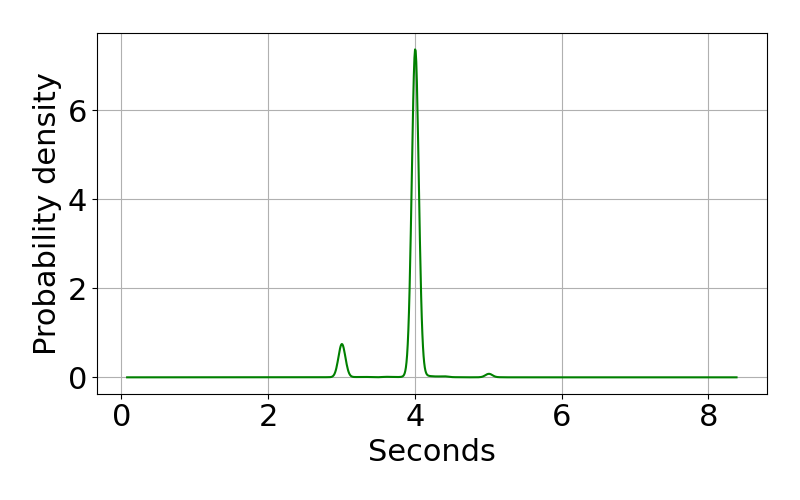}
        \caption{Pdf: validation interarrival time}
        \label{fig:Default behavior: Probability Density function}
    \end{subfigure}%
    \begin{subfigure}[t]{0.5\columnwidth}
        \centering
        \includegraphics[height=1.15in]{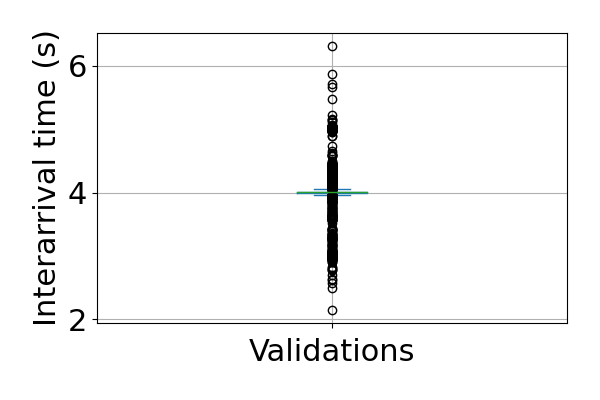}
        \caption{Validation interarrival time}
        \label{fig:Default behavior: whisker plot}
    \end{subfigure}
    \newline
    \begin{subfigure}[t]{\columnwidth}
        \centering
        \includegraphics[width=\linewidth]{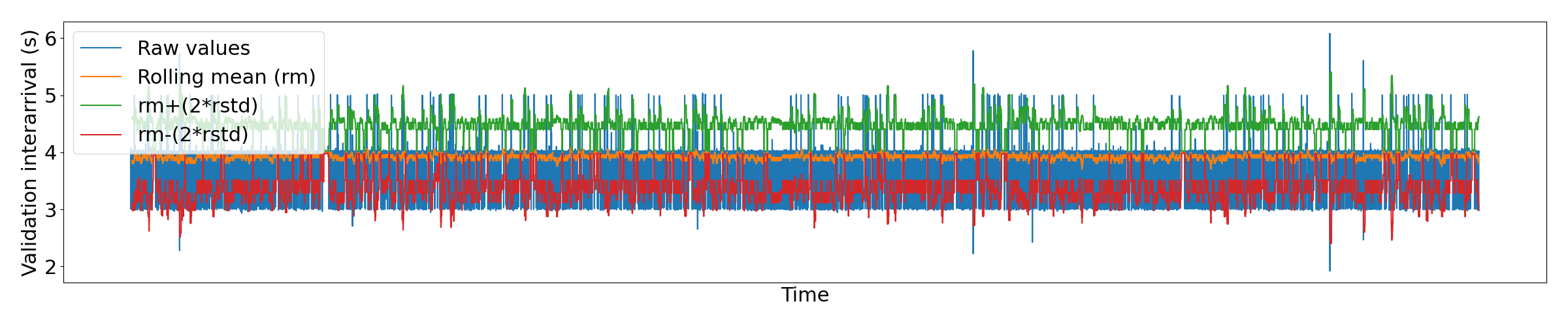}
        \caption{Time series: validation interarrival time}
        \label{fig:Production-default}
    \end{subfigure}
    \caption{Typical validation interarrival time on XRPL \textit{livenet}}
    \label{fig:typical-xrpl-livenet}
\end{figure}

\begin{figure*}[t!]
    \centering
    \begin{subfigure}[t]{0.5\textwidth}
        \centering
        \includegraphics[width=\linewidth]{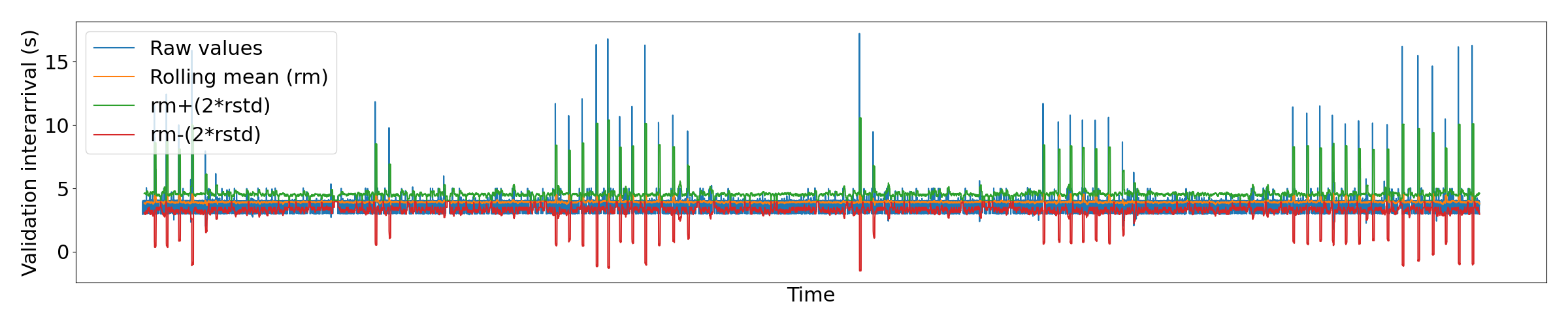}
        \caption{"Intermittent" behavior}
        \label{fig:Production-intermittent}
    \end{subfigure}%
    \begin{subfigure}[t]{0.5\textwidth}   
        \includegraphics[width=\linewidth]{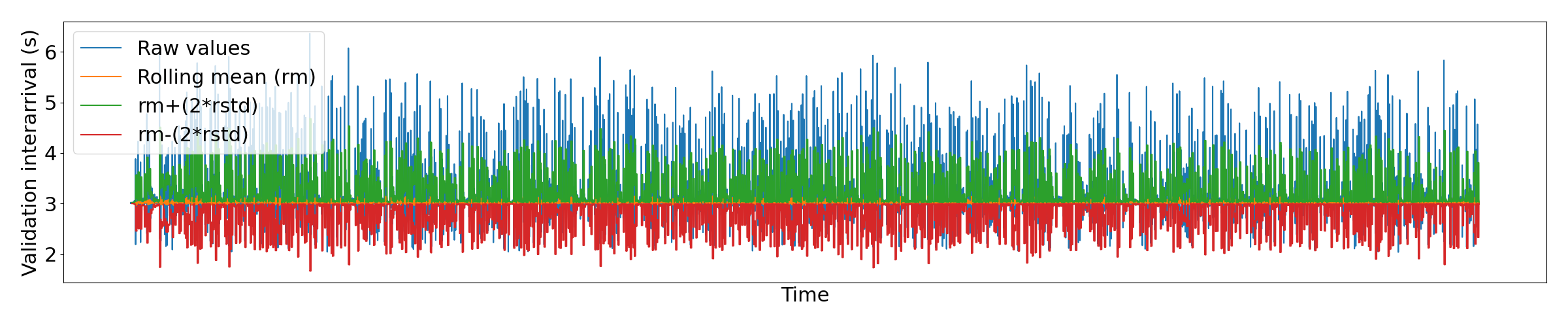}
        \caption{"Atypical" behavior}
        \label{fig:Production-atypical}
    \end{subfigure}
    \newline
    \begin{subfigure}[t]{0.5\textwidth}
        \centering
        \includegraphics[width=\linewidth]{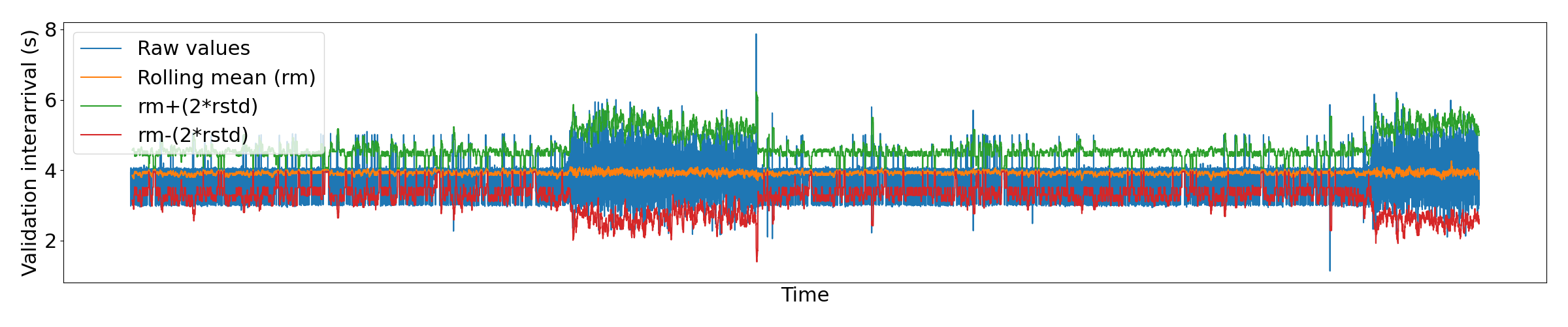}
        \caption{"One-off" behavior}
        \label{fig:Production-One-off}
    \end{subfigure}%
    \begin{subfigure}[t]{0.5\textwidth}   
        \includegraphics[width=\linewidth]{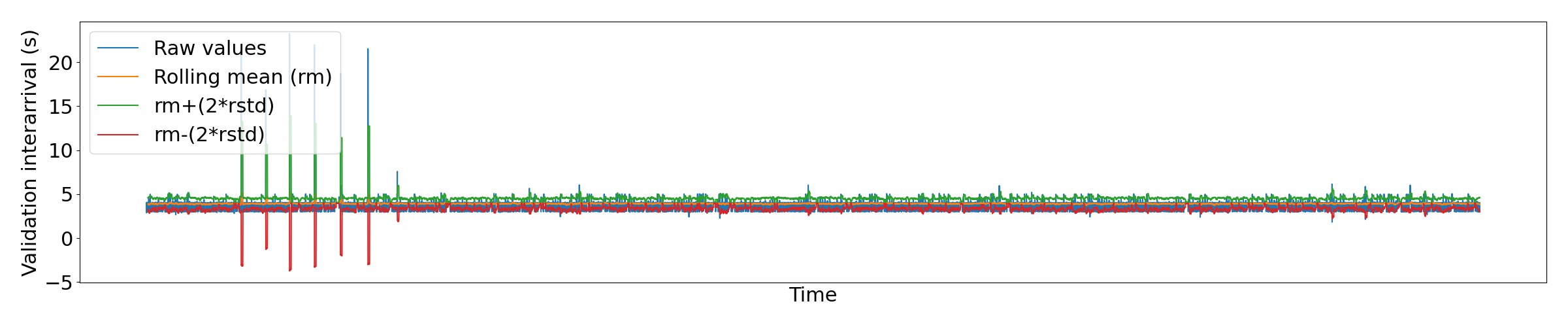}
        \caption{"One-off" behavior}
        \label{fig:Production-One-off2}
    \end{subfigure}
    \newline
    \begin{subfigure}[t]{0.5\textwidth}
        \centering
        \includegraphics[width=\linewidth]{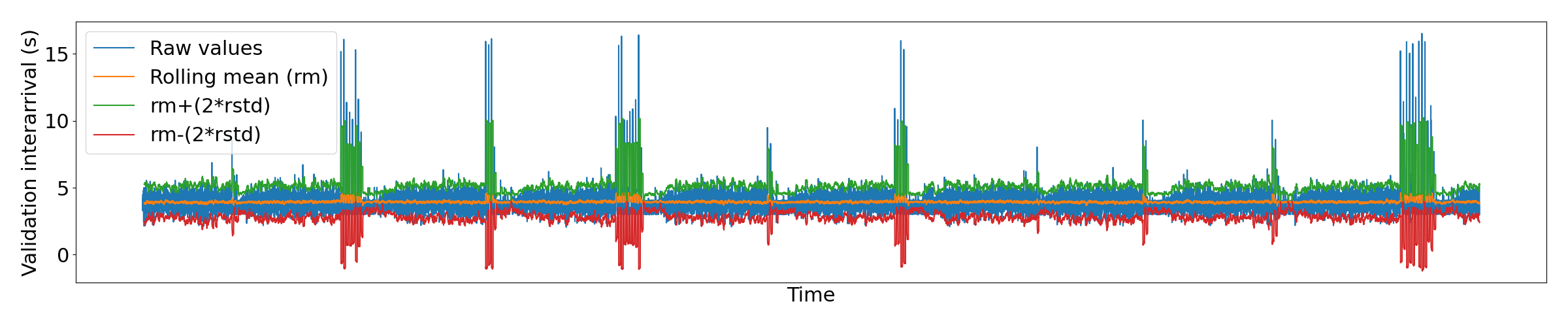}
        \caption{"Regular intervals" behavior}
        \label{fig:Production-regular3}
    \end{subfigure}%
    \begin{subfigure}[t]{0.5\textwidth}   
        \includegraphics[width=\linewidth]{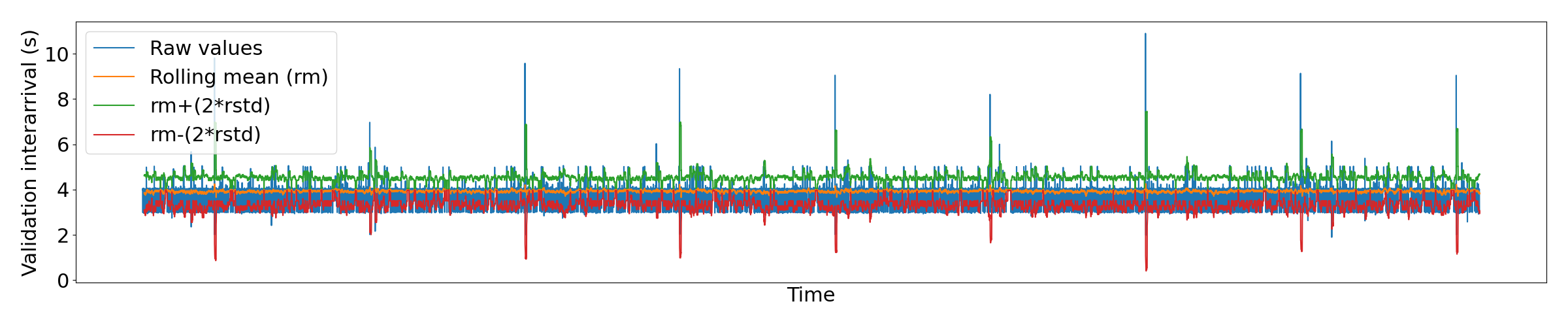}
        \caption{"Regular intervals" behavior}
        \label{fig:Production-regular2}
    \end{subfigure}
    \newline
    \begin{subfigure}[t]{\textwidth} 
        \centering
        \includegraphics[width=\linewidth]{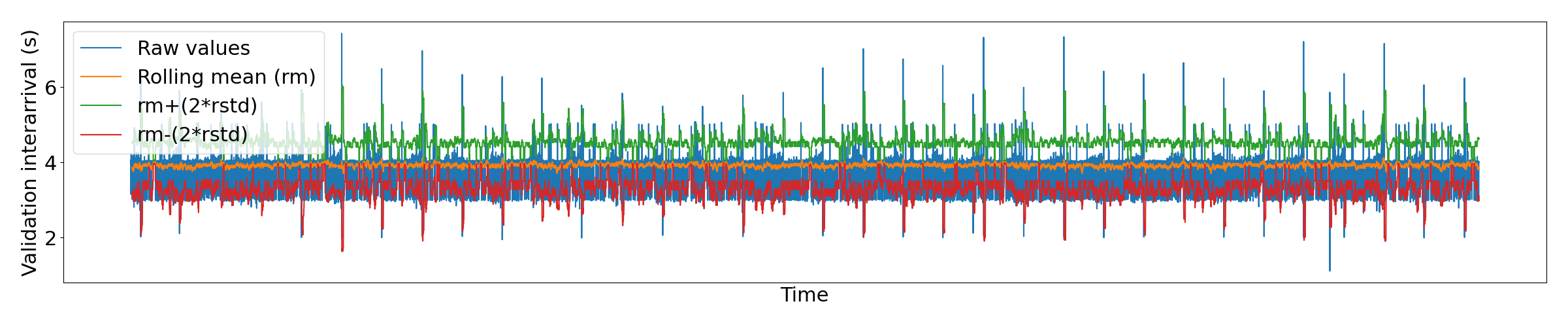}
        \caption{"Regular intervals" behavior.}
        \label{fig:Production-regular}
    \end{subfigure}
    \caption{Behaviors of validators on the XRPL \textit{livenet}, different from the typical behavior (time series)}
    \label{fig:diff-from-typical-behavior-xrpl-livenet}
\end{figure*}

\textbf{M3}: We notice that for the XRPL validators which we "listen" to, while the seemingly default behaviour would be to receive incoming validations spaced at between 3 to 5 seconds (with the mean approximately around 3.92s, median around 4s, quantile(0.25) around 3.98s and quantile(0.75) around 4.02s, as shown in Figures~\ref{fig:Production-default},~\ref{fig:Default behavior: Probability Density function},~\ref{fig:Default behavior: whisker plot}), there are validators which exhibit atypical, one-off, irregular or seemingly regular disruptions in the default pattern, of which we illustrate some cases in Figures~\ref{fig:Production-regular},~\ref{fig:Production-regular2},~\ref{fig:Production-regular3},~\ref{fig:Production-One-off},~\ref{fig:Production-One-off2},~\ref{fig:Production-intermittent},~\ref{fig:Production-atypical}.

\subsubsection{Private network of unmodified XRPL validators} On our dedicated testbed we set up a private network of unmodified, fully connected XRPL validators. From one of these, we record as above, the intervals between the first arrived unique validations from all other validator nodes, dropping duplicates. 

\textbf{M1}: Using \textit{RippledMon} and \textit{Grafana} we were able to compute the ratio of total validations in+out to ledgers created. It turns out that for the baseline topology from Figure~\ref{fig:experimental-topologies}(middle), there are on average 7.34 validations travelling in/out from an XRPL node to create one ledger, and a total number of 17845 validations over 2 hours.

\textbf{M2}: Over 10 minutes, \textit{Tshark} recorded 14420 packets. \textit{Vnstat} reported an average rate of 58kbit/s over 5 minutes.

\begin{figure}[t!]
    \centering
    \begin{subfigure}[t]{0.5\columnwidth}
        \centering
        \includegraphics[height=1.15in]{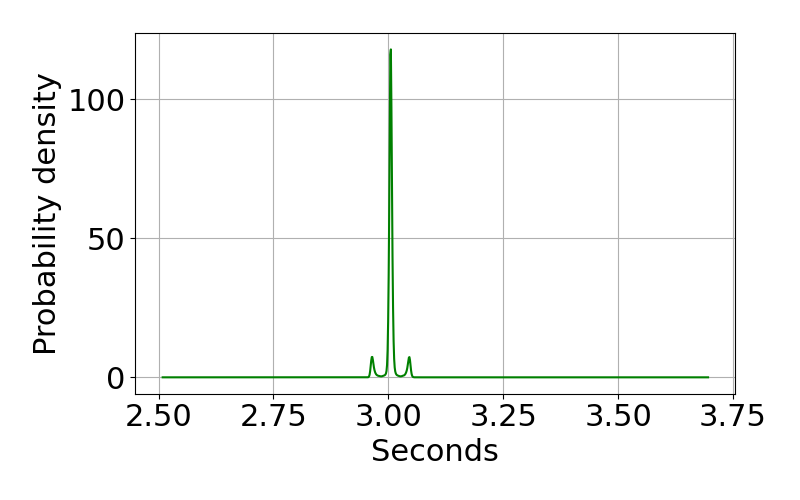}
        \caption{Pdf: validation interarrival time}
        \label{fig:Private-typical-prob-dist}
    \end{subfigure}%
    \begin{subfigure}[t]{0.5\columnwidth}
        \centering
        \includegraphics[height=1.15in]{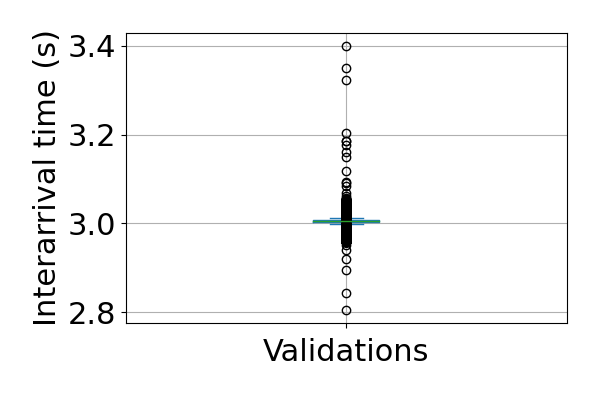}
        \caption{Validation interarrival time}
        \label{fig:Private-typical-whisker plot}
    \end{subfigure}
    \newline
    \begin{subfigure}[t]{\columnwidth}
        \centering
        \includegraphics[width=\linewidth]{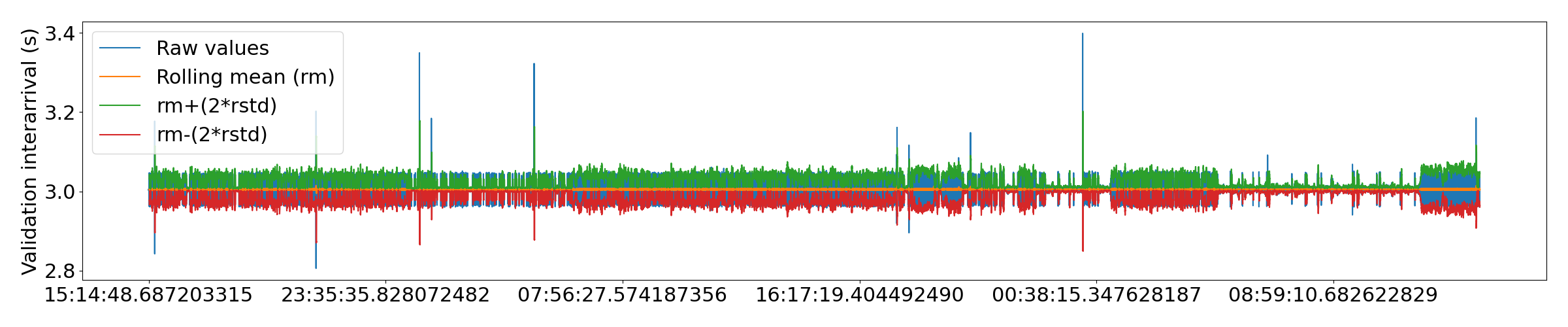}
        \caption{Time series: validation interarrival time}
        \label{fig:Private-typical-inter-arrivals}
    \end{subfigure}
    \caption{Validation interarrival time: \textit{baseline} (private XRPL)}
    \label{fig:Private-typical-times}
\end{figure}

\textbf{M3}: We notice a strong tendency for interarrival times spaced sharply at around 3 seconds with a mean, median, quantile(0.25) and quantile(0.75) all around 3.00s, as shown in Figures~\ref{fig:Private-typical-prob-dist},~\ref{fig:Private-typical-whisker plot} and \ref{fig:Private-typical-times}. These figures will be used later to compare the different models.

\subsubsection{The "polling" model} This was our first validation dissemination model, mostly to see how the XRPL and NDN would work out together. We carried out the experiments concerning the polling model on the \textit{triangle} topology.

\begin{figure}[htbp!]
    \centering
    \begin{subfigure}[t]{0.5\columnwidth}
        \centering
        \includegraphics[height=1.15in]{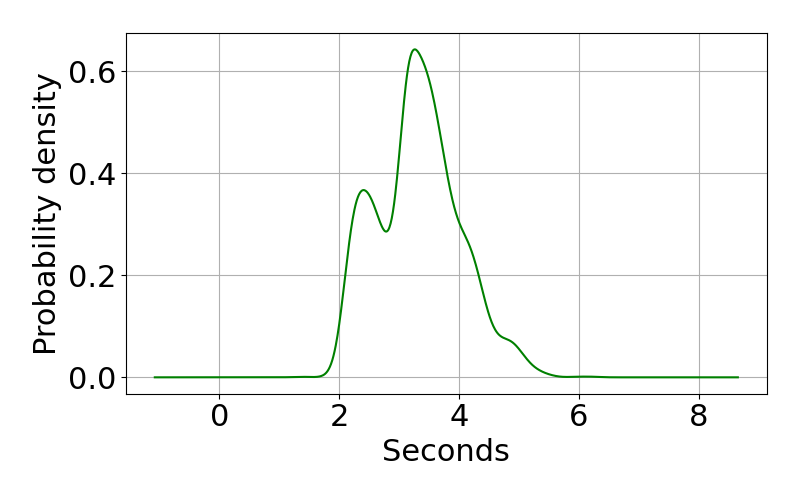}
        \caption{Pdf: validation interarrival time}
        \label{fig:Poll-prob-dist}
    \end{subfigure}%
    \begin{subfigure}[t]{0.5\columnwidth}
        \centering
        \includegraphics[height=1.15in]{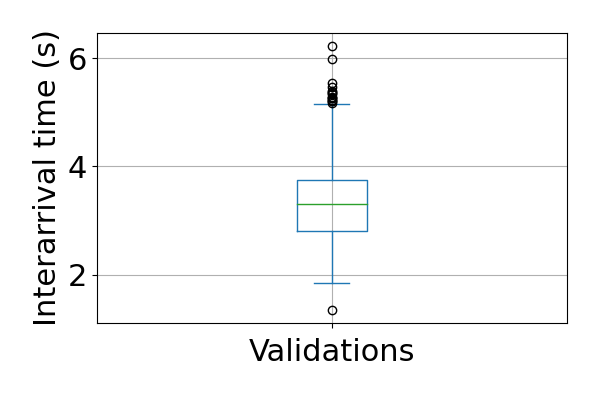}
        \caption{Validation interarrival time}
        \label{fig:Poll-whisker plot}
    \end{subfigure}
    \newline
        \begin{subfigure}[t]{\columnwidth}
        \centering
        \includegraphics[width=\linewidth]{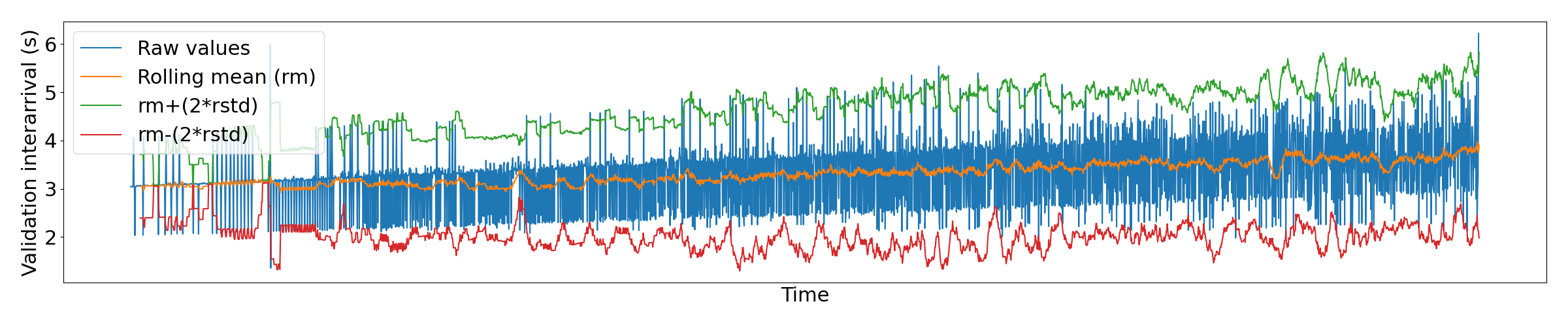}
        \caption{Time series: validation interarrival time}
        \label{fig:Poll-typical-val-inter}
    \end{subfigure}
    \caption{Validation interarrival time: \textit{Polling(tri)} model}
    \label{fig:Poll-typical-behavior}
\end{figure}

\textbf{M3}: The inter-arrival times from Figure~\ref{fig:Poll-typical-behavior} are generally not better than the baseline. This, together with the high number of messages incurred at NDN level by the continuous polling make this set-up unfeasible for a real-life usage.

Because this model performed worse than the \textit{baseline} and the \textit{Piggyback} model (described below) concerning validation interarrival times (Table~\ref{tab:summary}), we didn't collect any further metrics (M1 and M2). However, this model could be further improved to use for example adaptive polling intervals.

\subsubsection{The "announce-pull" model} We experimented with both the \textit{star} and \textit{triangle} topologies. This was the second model we experimented with, as an improvement over the first one. 

\begin{figure}[htbp!]
    \centering
    \begin{subfigure}[t]{0.5\columnwidth}
        \centering
        \includegraphics[height=1.15in]{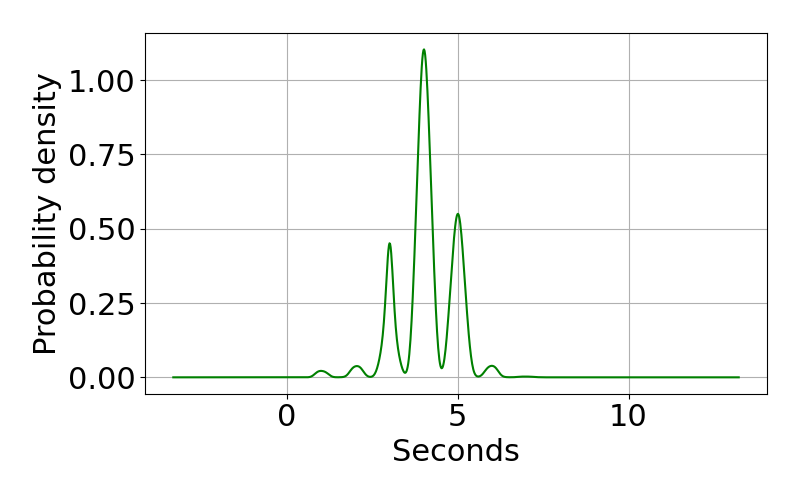}
        \caption{Pdf: validation interarrival time}
        \label{fig:Ann-Pull-triangle-prob-dist}
    \end{subfigure}%
    \begin{subfigure}[t]{0.5\columnwidth}
        \centering
        \includegraphics[height=1.15in]{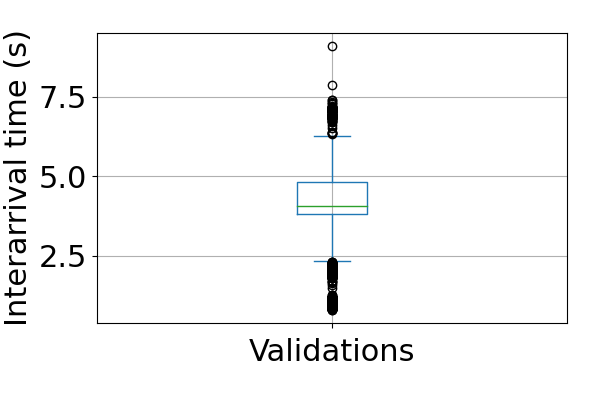}
        \caption{Validation interarrival time}
        \label{fig:Ann-Pull-triangle-whisker plot}
    \end{subfigure}
    \newline
    \begin{subfigure}[t]{\columnwidth}
        \centering
        \includegraphics[width=\linewidth]{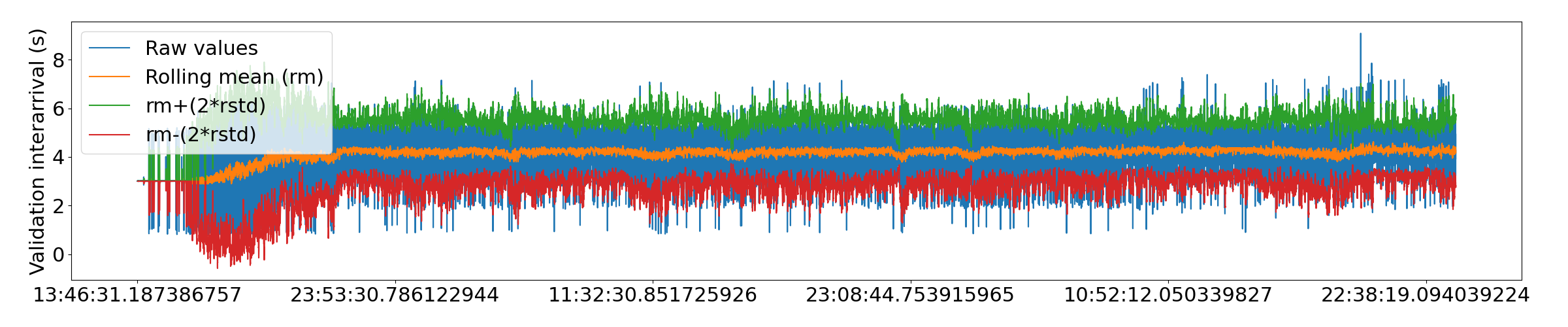}
        \caption{Time series: validation interarrival time}
        \label{fig:Ann-Pull-triangle-vals-interarr}
    \end{subfigure}    
    \caption{Validation interarrival time: \textit{Announce-Pull(tri)} model}
    \label{fig:Ann-pull-triangle-typical-behavior}
\end{figure}

On the \textit{triangle} topology, overall, this model showed a more stable behavior, however without getting close to the baseline concerning M3 (\(rm\) and \(rSTD\)), as shown in Figure~\ref{fig:Ann-pull-triangle-typical-behavior}.

\begin{figure}[t!]
    \centering
    \begin{subfigure}[t]{0.5\columnwidth}
        \centering
        \includegraphics[height=1.15in]{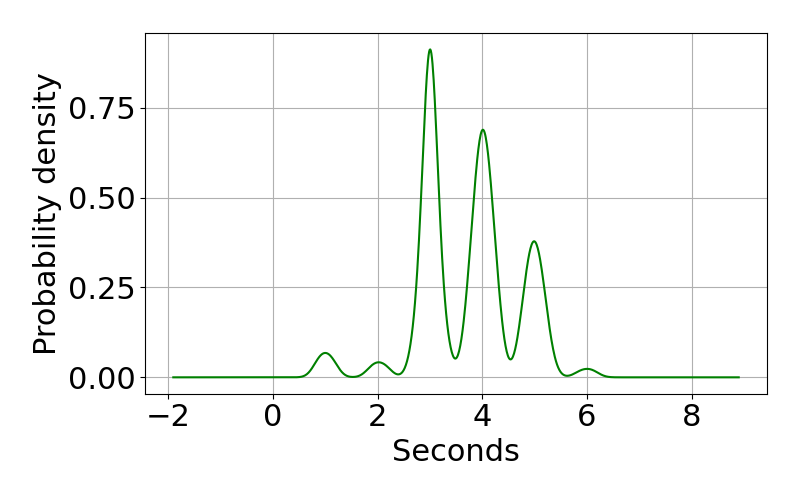}
        \caption{Pdf: validation interarrival time}
        \label{fig:Ann-Pull-prob-dist-star}
    \end{subfigure}%
    \begin{subfigure}[t]{0.5\columnwidth}
        \centering
        \includegraphics[height=1.15in]{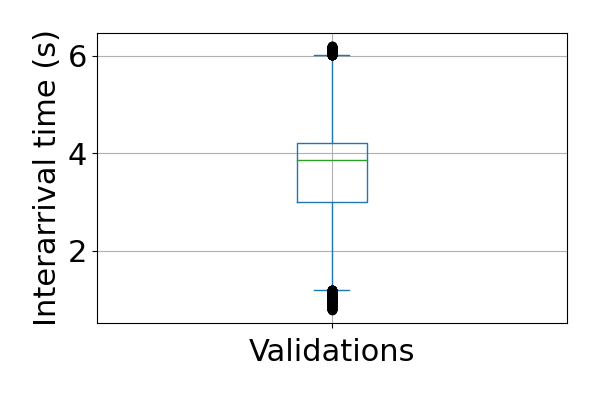}
        \caption{Validation interarrival time}
        \label{fig:Ann-Pull-whisker plot-star}
    \end{subfigure}
    \newline
    \begin{subfigure}[t]{\columnwidth}
        \centering
        \includegraphics[width=\linewidth]{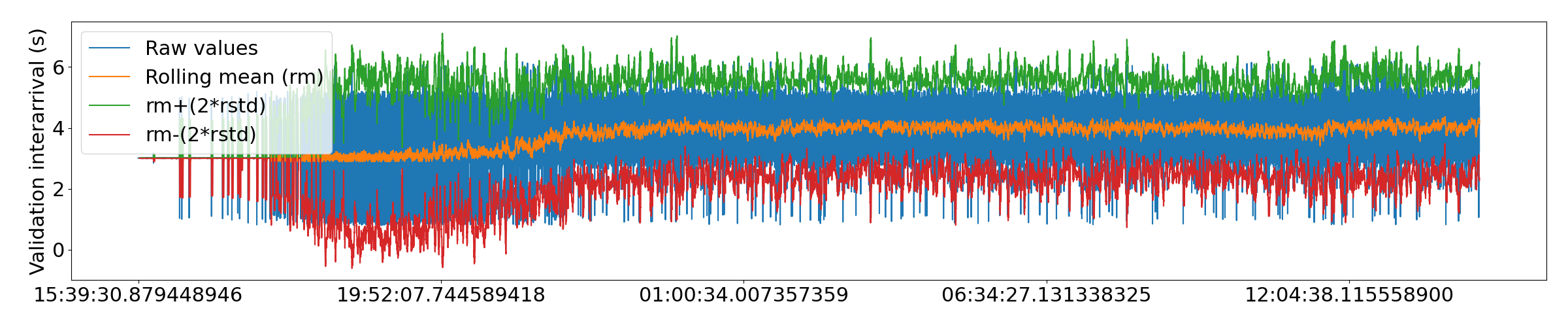}
        \caption{Time series: validation interarrival time}
        \label{fig:Ann-Pull-vals-inter-star}
    \end{subfigure}
    \caption{Validation interarrival time: \textit{Announce-Pull(star)}}
    \label{fig:Ann-pull-star-typical-behavior}
\end{figure}

On the \textit{star} topology, we present our results in Figure~\ref{fig:Ann-pull-star-typical-behavior}.

\begin{figure}[ht]
\begin{center}
    \includegraphics[width=1\columnwidth]{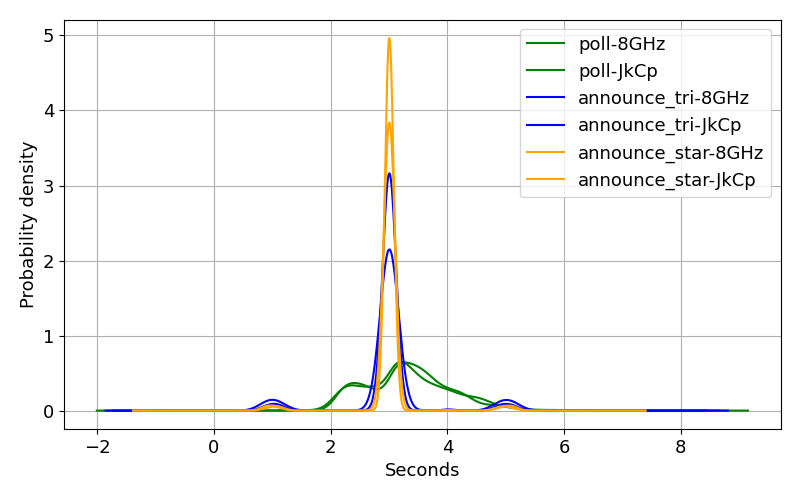}
    \caption{Pdf overlap for validations received at node1 from remote nodes for \textit{Polling(tri)} and \textit{Announce-pull(tri/star)} models}
    \label{fig:Poll-Ann-pull-overlap}
\end{center}
\end{figure}

In order to be able to perform a better comparison, in Figure~\ref{fig:Poll-Ann-pull-overlap} we present an overlap of the results on the 2 models.

From Figure~\ref{fig:Poll-Ann-pull-overlap} it follows that concerning M3, the \textit{announce-pull(star)} topology performed better compared to \textit{polling}, however because the \textit{announce-pull} model also performed worse than the \textit{baseline} and than the \textit{Piggyback} model concerning validation interarrival times (Table~\ref{tab:summary}), we didn't collect any further metrics (M1 and M2).

\subsubsection{The "advanced-request" model} 
\begin{figure}[t!]
    \centering
    \begin{subfigure}[t]{0.5\columnwidth}
        \centering
        \includegraphics[height=1.15in]{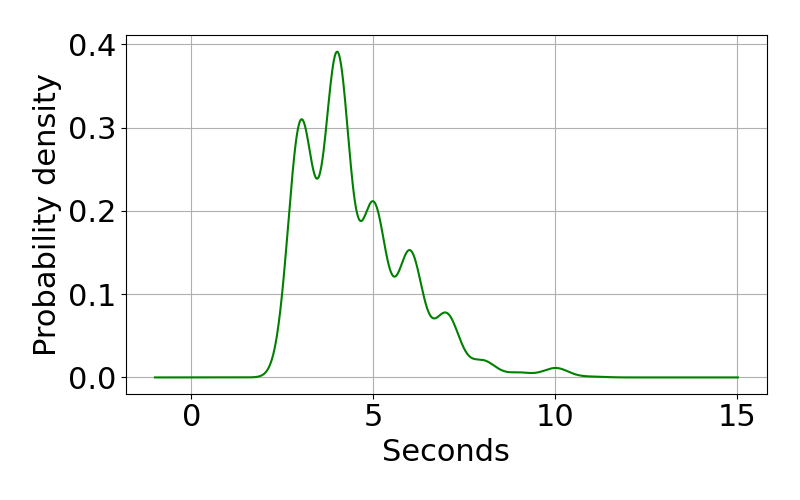}
        \caption{Pdf: validation interarrival time}
        \label{fig:Adv-req-triangle-prob-dist}
    \end{subfigure}%
    \begin{subfigure}[t]{0.5\columnwidth}
        \centering
        \includegraphics[height=1.15in]{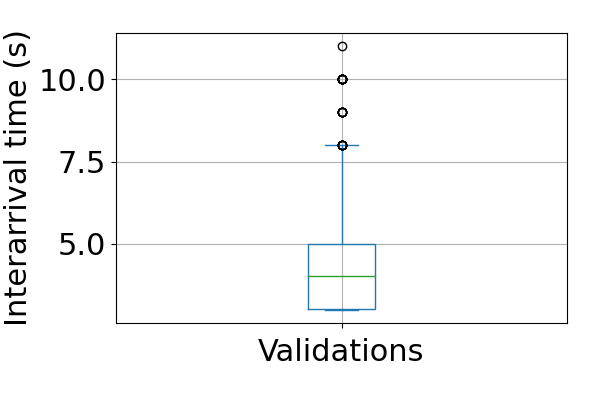}
        \caption{Validation interarrival time}
        \label{fig:Adv-req-triangle-whisker plot}
    \end{subfigure}
    \caption{Validation interarrival time: \textit{Advanced-request(tri)}}
    \label{fig:Adv-req-triangle-typical-behavior}
\end{figure}

The results obtained on the \textit{triangle} topology are presented in Figure~\ref{fig:Adv-req-triangle-typical-behavior}. This model was not investigated further as the performance under M3 was not satisfactory.

\subsubsection{The "Piggibacking on Interest" model}
\begin{figure}[t!]
    \centering
    \begin{subfigure}[t]{0.5\columnwidth}
        \centering
        \includegraphics[height=1.15in]{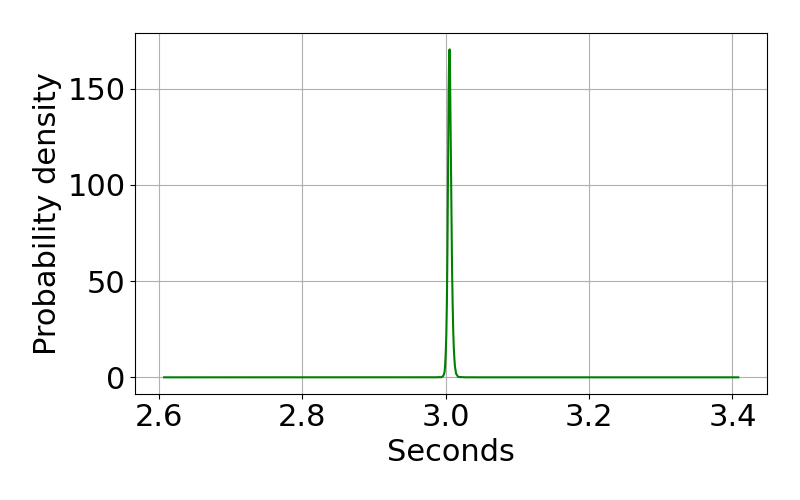}
        \caption{Pdf: validation interarrival time}
        \label{fig:Piggyback-triangle-prob-dist}
    \end{subfigure}%
    \begin{subfigure}[t]{0.5\columnwidth}
        \centering
        \includegraphics[height=1.15in]{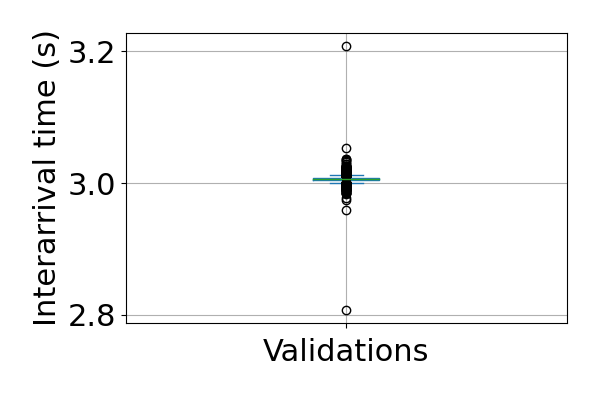}
        \caption{Validation interarrival time}
        \label{fig:Piggyback-triangle-whisker plot}
    \end{subfigure}
    \newline
    \begin{subfigure}[t]{\columnwidth}
        \centering
        \includegraphics[width=\linewidth]{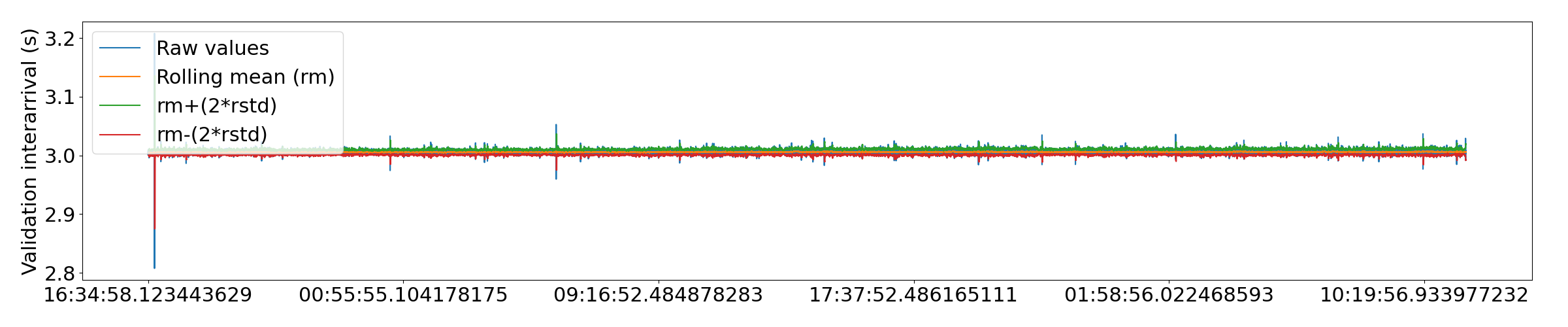}
        \caption{Time series: validation interarrival time}
        \label{fig:Piggyback-validations-inter-triangle}
    \end{subfigure}    
    \caption{Validation interarrival time: \textit{Piggyback(tri)} model}
    \label{fig:Piggyback-triangle-typical-behavior}
\end{figure}

: Under M1, there was a total number of 3 validations in+out of the XRPL node, per ledger created. Concerning M2, \textit{tshark} recorded over 10 minutes a number of 13713 packets, and \textit{vnstat} showed an average of 80kbit/s over 5 minutes. M3 is better than the baseline, as shown in Figure~\ref{fig:Piggyback-triangle-typical-behavior}.

The experiments were carried out on the "triangle" topology from Figure~\ref{fig:experimental-topologies}. The probability distribution plots show a slightly better performance of the piggyback model versus the baseline, while concerning the number of validations processed at the XRPL node, the piggyback model is clearly better with 3 validations/ledger versus 7.34, meaning that our model is 2.44 times better for \textit{this experimental setup}. 

All the above results are summed up in Table~\ref{tab:summary}.

\begin{table*}
\caption{Experiments summary}
\centering
\begin{tabular}{|l|c|ccc|cccccc|}
\hline
\multirow{2}{*}{\textbf{Model}}                                          & \multicolumn{1}{l|}{\multirow{2}{*}{\textbf{Topo}}} & \multicolumn{3}{c|}{\textbf{Val inter-arrival time}}                 & \multicolumn{1}{c|}{\textbf{XRP node load}} & \multicolumn{2}{c|}{\textbf{NIC load}}                                   & \multicolumn{3}{c|}{\textbf{Content Store (rates / min)}}                                                  \\ \cline{3-11} 
                                                                         & \multicolumn{1}{l|}{}                               & \multicolumn{1}{c|}{q(0.25)} & \multicolumn{1}{c|}{q(0.5)} & q(0.75) & \multicolumn{1}{c|}{vals in+out/ledger}     & \multicolumn{1}{c|}{avg bitrate (5min)} & \multicolumn{1}{c|}{pkt/10min} & \multicolumn{1}{c|}{misses}                      & \multicolumn{1}{c|}{hits} & entries                     \\ \hline
Baseline                                                                 & tri                                                 & 3.00                         & 3.00                        & 3.00    & \multicolumn{1}{c|}{7.34}                   & \multicolumn{1}{c|}{59kbit/s}           & \multicolumn{1}{c|}{14420}     & \multicolumn{1}{c|}{N/A}                         & \multicolumn{1}{c|}{N/A}  & N/A                         \\ \hline
Adv-req                                                                  & tri                                                 & 3.00                         & 4.00                        & 5.00    & \multicolumn{1}{c|}{not collected}          & \multicolumn{1}{c|}{20kbit/s}           & \multicolumn{1}{c|}{11800}     & \multicolumn{3}{c|}{not collected}                                                                         \\ \hline
Polling                                                                  & tri                                                 & 2.95                         & 3.48                        & 4.52    & \multicolumn{6}{c|}{not collected}                                                                                                                                                                                                  \\ \hline
\multirow{2}{*}{\begin{tabular}[c]{@{}l@{}}Announce\\ Pull\end{tabular}} & star                                                & 3.00                         & 3.86                        & 4.21    & \multicolumn{3}{c|}{not collected}                                                                                     & \multicolumn{1}{c|}{170-\textgreater{}790 (2h)}  & \multicolumn{1}{c|}{0}    & 887-\textgreater{}1520 (2h) \\
                                                                         & tri                                                 & 3.86                         & 4.07                        & 4.84    & \multicolumn{3}{c|}{not collected}                                                                                     & \multicolumn{1}{c|}{900-\textgreater{}1500 (2h)} & \multicolumn{1}{c|}{0}    & 190-785 (2h)                \\ \hline
Piggyback                                                                & tri                                                & 3.00                         & 3.00                        & 3.00    & \multicolumn{1}{c|}{3}                      & \multicolumn{1}{c|}{80kbit/s}           & \multicolumn{1}{c|}{13700}     & \multicolumn{1}{c|}{785 (flat)}                  & \multicolumn{1}{c|}{0}    & 65 (flat)                   \\ \hline
\end{tabular}
\label{tab:summary}
\end{table*}

According to these results, for the concrete case of XRPL validations, we find that the \textit{best suitable solution} is their encapsulation in Interest messages and dissemination with multicast. This approach uses few additional messages (the goal was to minimise the overall number of messages, and the ratio we obtained was 3 to 7 between our model and the baseline, respectively). This model improves significantly over the baseline as shown by comparing the interarrival times, while ensuring robust dissemination and low latency.


\section{Related Work}
\label{sec:relwork}


	Focuses on block propagation delay not flooding/bandwidth issue 

\subsection{Data synchronisation protocols}

Being content oriented, the NDN architecture suites for data synchronisation, which resulted in the development of different data synchronisation protocols, like \textit{Vectorsync \cite{vectorsync}, Chronosync \cite{chronosync}}, or \textit{Psync \cite{pSync}}. 

\textit{BlockNDN}~\cite{BlockNDN} employs the Chronosync protocol for block broadcasting on a Bitcoin-like blockchain and \textit{State Vector Sync} \cite{state_vector_sync} was also proposed to sync data between multiple nodes over NDN. However, in \cite{NDN-Ethereum}, it is argued that these solutions are not suitable for blockchain networks because, between others, the underlying protocol used was not originally designed for the byzantine environment characteristic to DLT. Moreover for XRPL, if the goal is to minimise the number of messages incurred by the communication, then the synchronisation protocols could introduce some additional synchronisation messages which could be counter-productive.

    
\subsection{Other approaches concerning DLTs and NDN} The problem of communication efficiency for blockchain networks at scale has lately received increased attention:

\textit{BoNDN}~\cite{BoNDN}, proposes tx dissemination for Bitcoin (BTC) through a push model over NDN interests, and a subscribe-push model for block propagation. The model proposed for tx dissemination is similar to our approach \textit{piggyback} where we obtained good results performance-wise, but it is challenged in \cite{NDN-Ethereum} for using multicast at NDN level - the authors argue that it is doubtful if in practice, the NDN nodes would enable multicast for the given data-labels. For XRPL which incurs a handful of known-in-advance, static list of validators, we find such a set-up not problematic.

\cite{NDN-Ethereum} proposes a design and implementation for propagating the ETH tx's and blocks over NDN. However, the design is focused on PoW blockchains, with a concrete case for ETH. We argue that the needs of consensus-validation based DLTs are fairly different from those of PoW DLTs, to require separate consideration. For example the size of the consensus messages in XRPL (proposals and validations) is much smaller than the size of ETH blocks, and XRPL uses the concept of UNLs where, strictly from a consensus perspective, a validator needs only receive messages from those other validators in its defined UNL. Moreover on NDN, the data can be signed and dated by the producer, which in XRPL case is already known (UNL validators are known), making some types of attacks discussed in this paper not applicable for our work on \system.
The authors dismiss NDN data sync models for various reasons, including security. While for XRPL's validations for example, the sync vector can be easily constructed, we agree that these models could add unnecessary traffic hindering scalability, and also, that they were not designed with Bizantine failures in mind.

The previous work also proposes an \textit{announce-pull} model for both tx and block propagation, arguing that this can benefit from in-network caching and multicasting to avoid redundant traffic. For the case of XRPL, because a \textit{Consumer} knows in advance the identity of the originating \textit{Producer} (a validator on its UNL), and because the interval between new validations is somewhat predictable (3-5 seconds in real-life) this model can be simplified to consider the announce already made, and issue pull requests in advance. Moreover, the authors use the ETH P2P overlay to broadcast the creation of a new block and then NDN to pull the block after learning about it. The problem on XRPL is fundamentally different and consists of a very large number of messages - a result of the flood mechanism when the network is scaled up. For XRPL, this number needs to be minimised. As such in this work we have been searched for and proposed a paradigm suitable to XRPL.




\section{Conclusions and Future Work}
\label{sec:conclusion}

The XRP ledger has a flooding mechanism that lacked peer-reviewed research, with most works having focused on other aspects like the consensus mechanism. Therefore, in this paper we investigate how flooding can be optimised using NDN. This is a promising overlay candidate because of its well researched and optimised caching and flooding mechanisms.

In this work, we i) proposed multiple mapping models for the transmission of the consensus-related messages, and ii) investigated the advantages and disadvantages of each of these models according to the specific needs of the blockchains using a consensus-validation system.

XRPL \textit{Proposal} messages share similar characteristics to \textit{validations} and could use the same dissemination model. Because of specific use cases such as trading or high frequency trending, to ensure that \textit{transactions} propagate as fast as possible to be included in the earliest ledger, the transactions could also use piggybacking on Interest messages. This approach however, might be subject to poisoning attacks and thus require additional mitigation measures, such as in-flight transaction verification, auditing, or node scoring.

We used a real testbed and the original XRPL code which required a significant effort to integrate NDN. The code is open source and can be found at the location below: \textit{https://github.com/FlavScheidt/sntrippled}.

\textit{Limitations:} Experimentation was limited to the scenarios and topologies on which we report. We plan to address larger and  more life-like topologies in order to assess the security and the robustness of our solution versus flooding. The analysis of different behaviors of the validators on the default UNL on the live XRP network with machine learning tools is also on our roadmap.

The XRPL consensus leverages the concept of UNLs where a validator may want to be interested to hear only validations from nodes on its own UNL. Currently in production, only two largely overlapping UNLs co-exist. Using more UNLs in production will not impact NDN traffic on any local NDN node since the local NDN nodes can be independently set to also relay any other intended traffic at NDN level.

\section*{Acknowledgment}
%
This work is supported by the Luxembourg National Research Fund through grant PRIDE15/10621687/SPsquared. In addition, we thankfully acknowledge the support from the RIPPLE University Blockchain Research Initiative (UBRI) for our research.  


\bibliographystyle{style/IEEEtran}
\bibliography{bib/ebpf}

\end{document}